\documentclass[prd,aps,amsfonts,superscriptaddress,nofootinbib,longbibliography,notitlepage,twocolumn]{revtex4-1}
\usepackage{graphicx}
\usepackage[dvipsnames]{xcolor}
\usepackage{rotating}
\usepackage{amsmath,amssymb,amsfonts,dsfont,mathtools}
\usepackage{amsthm}
\usepackage{array}

\usepackage{comment}

\usepackage[colorlinks=true, urlcolor=violet, linkcolor=blue, citecolor=red, hyperindex=true, linktocpage=true]{hyperref}

\numberwithin{thm}{section}

\makeatletter
\renewcommand{\p@subsection}{}
\renewcommand{\p@subsubsection}{}
\makeatother

\usepackage[normalem]{ulem} 

\newcommand{\p}{\partial}

\DeclarePairedDelimiter{\abs}{\lvert}{\rvert}

\makeatletter
\DeclareRobustCommand{\cev}[1]{%
  \mathpalette\do@cev{#1}%
}
\newcommand{\do@cev}[2]{%
  \fix@cev{#1}{+}%
  \reflectbox{$\m@th#1\vec{\reflectbox{$\fix@cev{#1}{-}\m@th#1#2\fix@cev{#1}{+}$}}$}%
  \fix@cev{#1}{-}%
}
\newcommand{\fix@cev}[2]{%
  \ifx#1\displaystyle
    \mkern#23mu
  \else
    \ifx#1\textstyle
      \mkern#23mu
    \else
      \ifx#1\scriptstyle
        \mkern#22mu
      \else
        \mkern#22mu
      \fi
    \fi
  \fi
}

\makeatother

\newcolumntype{P}[1]{>{\centering\arraybackslash}p{#1}}

\begin{document}

\title{Fracton hydrodynamics without time-reversal symmetry}

\author{Jinkang Guo}
\affiliation{Department of Physics and Center for Theory of Quantum Matter, University of Colorado, Boulder CO 80309, USA}

\author{Paolo Glorioso}
\email{paolog@stanford.edu}
\affiliation{Department of Physics, Stanford University, Stanford, CA 94305, USA}

\author{Andrew Lucas}
\email{andrew.j.lucas@colorado.edu}
\affiliation{Department of Physics and Center for Theory of Quantum Matter, University of Colorado, Boulder CO 80309, USA}

\begin{abstract}
We present an effective field theory for the nonlinear fluctuating hydrodynamics of a single conserved charge with or without time-reversal symmetry, based on the Martin-Siggia-Rose formalism.  Applying this formalism to fluids with only charge and multipole conservation, and with broken time-reversal symmetry, we predict infinitely many new dynamical universality classes, including some with arbitrarily large upper critical dimensions. Using large scale simulations of classical Markov chains, we find numerical evidence for a breakdown of hydrodynamics in quadrupole-conserving models with broken time-reversal symmetry in one spatial dimension.
\end{abstract}

\date{\today}

\maketitle

\section{Introduction}
In the past few years, infinitely many universality classes of hydrodynamics have been discovered \cite{fractonhydro,morningstar,knap2020,zhang2020,IaconisVijayNandkishore,iaconis2021,doshi,knap2021,Glorioso:2021bif,Grosvenor:2021rrt,osborne,Burchards:2022lqr,hart2021hidden,sala2021dynamics}, with exotic conservation laws such as the conservation of multipole charges or charges along sub-dimensional manifolds.  Dubbed ``fracton fluids", as such universality classes describe the thermalization of generic models of interacting fractons (particles with mobility constraints) \cite{vijay2015new,vijay2016fracton,pretko2017emergent,pretko2017generalized,pretko2017subdimensional,Pretko2018,gromov2018towards,slagle2017quantum,slagle2018symmetric,Seiberg_2020,10.21468/SciPostPhys.10.2.027}, a careful study of these new hydrodynamic universality classes is likely to give valuable insight into the foundational underpinnings of hydrodynamics as an effective field theory (EFT) \cite{Crossley:2015evo,Haehl:2015foa,Jensen:2017kzi}, especially in non-thermal systems with unusual symmetries.

In this letter, we find new universality classes of fracton hydrodynamics with broken time reversal symmetry.  To understand why this construction is subtle, let us consider the simplest fracton fluid: a 1d system with charge and dipole symmetry \cite{fractonhydro,morningstar,knap2020,zhang2020}, which can be experimentally realized in tilted optical lattices \cite{Guardado_Sanchez_2020}.  Letting $\rho$ denote the density of conserved charge, one finds that dipole conservation $\partial_t \int \mathrm{d}x \; x\rho = 0$ mandates \begin{equation}
    \partial_t \rho + \partial_x^2 J_{xx} = 0.
\end{equation}
With time reversal symmetry, 
\begin{equation} \label{jxx}
     J_{xx} = D\partial_x^2 \rho + \cdots 
\end{equation}
is necessary, where the dots denote subleading terms in derivative expansion.  Thus far, this result is justified using effective field theory methods based on coupling this fluid to background (mixed-rank) gauge fields \cite{fractonhydro}; a more straightforward argument is to note that $J_{xx}$ is time reversal odd, and thus only derivatives of $\rho$ can appear in $J_{xx}$, since time reversal is not broken within ideal hydrodynamics.  When time reversal symmetry is broken, is it possible to write $J_{xx} = -D^\prime \rho + \cdots$? 

Our purpose in this letter is to give a systematic and highly generalizable framework capable of answering this question (negatively).  We will develop a systematic effective field theory framework for studying hydrodynamics of non-thermal systems, with or without time reversal symmetry.  Studying many different examples of fracton fluids without time reversal symmetry, we will discover an infinite new family of dynamical universality classes, which generalize Kardar-Parisi-Zhang (KPZ) \cite{kpz,Spohn_2014,PhysRevE.90.012124,gloriosoprl} and multipolar extensions thereof \cite{Glorioso:2021bif}.

\section{Effective field theory}
We first develop a user-friendly EFT for a non-thermal fluid (one in which energy is not conserved, and temperature is not well-defined).  We focus on systems with a single conserved charge with density $\rho$, which is a scalar under rotations, inversions and time reversal; generalizations will appear elsewhere.  We assume that dynamics is local in space, ergodic, and that there exists a steady state probability distribution on the classical state space (or quantum density matrix) invariant under the microscopic dynamics.

For pedagogical purposes, consider nonlinear fluctuating hydrodynamics from a traditional perspective via classical stochastic differential equations \cite{gardiner}. (Note that our eventual EFT will also describe the hydrodynamics of microscopically quantum systems.)  It is useful (for now) to think of $\rho_x$ as the discretization of a continuum function $\rho(x)$ onto some $d$-dimensional lattice.  We write
    \begin{eqnarray}\label{langevin}
    \frac{\mathrm{d}\rho_x}{\mathrm{d}t} = F_x(\boldsymbol{\rho}) + \zeta_x(t),
    \end{eqnarray}
    where $F_x$ is some nonlinear function of $\rho$s on nearby lattice sites, consistent with all necessary symmetries, and $\zeta_x(t)$ corresponds to stochastic fluctuations. (\ref{langevin}) is in the Ito interpretation.  Eventually, we'll want a rulebook for how to calculate $F_x$ and the statistics of $\zeta_x$.  For now, assume that the noise is white, with zero-mean and \begin{eqnarray}\label{noise1}
            \langle \zeta_x(t) \zeta_{x^\prime}(t^\prime)\rangle = \epsilon Q_{xx^\prime}(\boldsymbol{\rho}) \delta(t-t^\prime),
    \end{eqnarray}
    with $\epsilon$ a perturbatively small ``bookkeeping" parameter, and $Q_{ij}$ symmetric and positive semidefinite.   It will be useful to replace (\ref{langevin}) by the equivalent Fokker-Planck equation for $P(\boldsymbol{\rho}, t)$,  
    \begin{align} \label{fokkerplanck}
	    \frac{\partial P}{\partial t}=\frac{\partial}{\partial \rho_{x}}\left[-F_{x}(\boldsymbol{\rho})P+\frac{\epsilon}{2}\frac{\partial }{\partial \rho_{x^\prime}}\left( Q_{xx^\prime}(\boldsymbol{\rho})  P\right)\right] .
    \end{align}
where summation over repeated indices is understood.

Now we bring in our first key assumption: the existence of a steady state distribution \begin{eqnarray} \label{ss}
        P_{\mathrm{eq}}(\boldsymbol{\rho}) \propto \exp[-\Phi(\boldsymbol{\rho})/\epsilon],
\end{eqnarray}
If $\epsilon \rightarrow 0$, this distribution becomes tightly peaked near minima of $\Phi$ at small $\epsilon$.  This limit is both technically convenient and physically sensible: on very long scales, a fluid should be approximately described by noise-free partial differential equations (e.g. Fick's Law).  Combining (\ref{fokkerplanck}) and (\ref{ss}) we conclude that  \cite{maier}  \begin{align}\label{Hamiltonian}
		-\mathrm{i} H(-\mathrm{i} \boldsymbol{\mu/}\epsilon, \boldsymbol{\rho}) &\equiv \frac{1}{\epsilon}\left( - F_x \mu_x + \frac{1}{2}Q_{xx^\prime}\mu_x \mu_{x^\prime}\right) \notag \\
		&= 0 + \mathrm{O}(\epsilon^0),  
	\end{align}
	where we have defined
	\begin{eqnarray}
	         \mu_x \equiv - \frac{\partial \Phi}{\partial \rho_x}.
	\end{eqnarray}
Already, we can see sharp connections to thermodynamics and statistical mechanics: $\Phi$ plays the role of entropy $S$, the thermodynamic potential in the microcanonical ensemble, while $\mu$ is the chemical potential conjugate to $\rho_x$.  This emergent thermodynamics does not require finite temperature, energy conservation, or time reversal symmetry.  Moreover, the noise variance $Q_{xx^\prime}$ is not arbitrary: (\ref{Hamiltonian}) mandates a fluctuation-dissipation theorem \cite{kwon,pingao} relating $Q_{xx^\prime}$ to $F_x$; the consequences of this will be especially clear in the EFT language.

In (\ref{Hamiltonian}), we also defined a function ${H(-\mathrm{i}\boldsymbol{\mu}/\epsilon, \boldsymbol{\rho})}$. We will now show that it can be interpreted as a `Hamiltonian'.
The path integral of the system described by (\ref{langevin}) is given by the Martin-Siggia-Rose method \cite{msr}:
    \begin{eqnarray}
		Z = \int \mathrm{D} \boldsymbol{\rho} \mathrm{D} \boldsymbol{\zeta}\  \delta\left(\partial_{t}\boldsymbol{\rho} - \boldsymbol{F}(\boldsymbol{\rho}) + \boldsymbol{\zeta} \right) \mathrm{e}^{-\int \mathrm{d} t\ \frac{1}{2\epsilon} \boldsymbol{\zeta} \boldsymbol{Q}^{-1} \boldsymbol{\zeta}},
	\end{eqnarray}
which is equivalent to 
\begin{align}
		Z & = \int \mathrm{D} \boldsymbol{\rho} \mathrm{D} \boldsymbol{\pi} \mathrm{D} \boldsymbol{\zeta}\ \mathrm{e}^{\mathrm{i} \int \mathrm{d} t\left(\boldsymbol{\pi} \partial_{t}\boldsymbol{\rho} - \boldsymbol{F}(\boldsymbol{\rho}) \cdot \boldsymbol{\pi}+ \frac{\mathrm{i}}{2\epsilon} \boldsymbol{\zeta} \boldsymbol{Q}^{-1} \boldsymbol{\zeta}+\boldsymbol{\zeta} \boldsymbol{\pi} \right)} \notag \\
		& = \int \mathrm{D} \boldsymbol{\rho} \mathrm{D} \boldsymbol{\pi}\ \mathrm{e}^{\mathrm{i} \int \mathrm{d} t \mathcal{L}} .
\end{align}
In the last equation we get the effective Lagrangian
\begin{align}\label{lagrangian}
		\mathcal{L} & = \boldsymbol{\pi} \partial_{t}\boldsymbol{\rho} - \boldsymbol{F}(\boldsymbol{\rho}) \boldsymbol{\pi}+ \frac{\mathrm{i}\epsilon}{2} \boldsymbol{\pi} \boldsymbol{Q} \boldsymbol{\pi}  = \boldsymbol{\pi} \partial_{t}\boldsymbol{\rho} - H(\boldsymbol{\pi}, \boldsymbol{\rho}).
\end{align}
Note that ${H(\boldsymbol{\pi}=-i \boldsymbol{\mu}/\epsilon, \boldsymbol{p})}$ is simply (\ref{Hamiltonian}) up to ${\mathrm{O}(\epsilon)}$.

From now on, we replace ${\rho_x}$ with its continuum limit ${\rho(x)}$. ${F_{x}}$, ${Q_{xx'}}$ and ${P(\boldsymbol{\rho},t)}$ then become functionals of ${\rho(x)}$.
The Hamiltonian in the continuum limit is 
\begin{align}\label{lagrangiandensity}
    H = \int \mathrm{d}x\left( F(x,\rho) \pi(x) - \frac{i\epsilon}{2} \pi(x) Q(x,\rho) \pi(x)\right),
\end{align}
where $F$ and $Q$ can include spatial derivatives acting on $\rho$ and/or $\pi$.

There are three important types of symmetries and constraints we wish to impose within the EFT.  We will begin by discussing them from a purely EFT perspective.

\textbf{Charge/multipole conservation:} 
For any integrable function ${f(x)}$, we define a "multipolar" charge as 
    \begin{eqnarray}
		Q_{f}:=\int \mathrm{d}^{d} x\ f\left(x\right) \rho(x).
	\end{eqnarray}
${Q_{f_{i}}}$ is conserved if the system is invariant under
    \begin{eqnarray} \label{cond2}
		\pi(x) \rightarrow \pi(x) + f(x) c(t),
	\end{eqnarray}
where ${c_{i}(t)}$ is an arbitrary function of time. Under this transformation, the action transforms as
    \begin{eqnarray}
		S \rightarrow S + \int \mathrm{d} t\ \mathrm{d}^{d} x\ f(x) c(t) \partial_{t}\rho(x).
	\end{eqnarray}
The invariance of the action gives
    \begin{eqnarray}
		\frac{\delta S}{\delta c(t)} = \frac{\mathrm{d}}{\mathrm{d} t} \int \mathrm{d}^{d} x f\left(x\right) \rho(x) = \frac{\mathrm{d}}{\mathrm{d} t} Q_{f} = 0.
	\end{eqnarray}
	
\textbf{Parity:} Under parity, ${x \rightarrow -x}$ and $\rho(x) \rightarrow \rho(-x)$. We further demand the canonical momentum ${\pi(x) \rightarrow \pi(-x)}$.
	
 \textbf{Time-reversal:}  Under time reversal, ${t \rightarrow -t}$ and $\rho(x,t) \rightarrow \rho(x,-t)$. For a general system that has time reversal symmetry and satisfy (\ref{cond1}), in order for the Lagrangian to be invariant under time reversal, the term ${\pi \partial_{t}\rho}$'s contribution to the action should remain the same. Under time reversal, ${\partial_{t}\rho \rightarrow -\partial_{t}\rho}$. If ${\pi \rightarrow -\pi}$ under time reversal, from the invariance of Hamiltonian, ${H(\pi,\rho) = H(-\pi,\rho)}$, we would find that the leading order of ${\pi}$ in the time-derivative free terms of Hamiltonian is ${H \sim \pi^{2}}$, which means the dynamics of the system is fully stochastic.

If we want a system whose dynamics is not fully stochastic, we have to change the behavior of ${\pi}$ under time reversal, namely ${\pi(x,t) \rightarrow -\pi(x,-t) + ig(x)}$, so now ${H(\pi,\rho) = H(-\pi+ ig,\rho)}$. From the above analysis, we konw that only when ${g \partial_{t}\rho}$ is a total derivative can the equations of motion be invariant. According to (\ref{Hamiltonian}), the Hamiltonian satisfies 
    \begin{eqnarray}\label{cond1}
		H(0, \rho) = H(-\mathrm{i} \mu/\epsilon, \rho) + \mathrm{O}(\epsilon^{0}) = 0.
	\end{eqnarray}
A natural choice is therefore
    \begin{eqnarray}\label{KMS}
		\pi(x,t) \rightarrow -\pi(x,-t) - i\mu(x)/\epsilon.
	\end{eqnarray}
	
Note that (\ref{KMS}) is a $\mathbb{Z}_2$ transformation reminiscent of the Kubo-Martin-Schwinger (KMS) symmetry used to implement time-reversal symmetry in dissipative thermal systems at temperture $T$.   It is consistent with the condition that two applications of the time reversal should return dynamical fields to their original values. (\ref{KMS}) is the unique kind of $\mathbb{Z}_2$ transformation on functions (also called an involution) not requiring an infinite order series in ${\pi}$. Since $\mu$ is a total derivative, assuming that $H$ is invariant under (\ref{KMS}), the change in the action is a total derivative: \begin{eqnarray} \label{Sshift}
        S \rightarrow S + \mathrm{i} \mathrm{\Delta}\Phi/\epsilon,
\end{eqnarray}
where $\mathrm{\Delta}\Phi$ denotes the difference in the thermodynamic potential $\Phi$ in the initial and final state. 

Remarkably, our EFT-based guess for how to implement time-reversal can also be justified \emph{microscopically}.  Assuming statistical time-translation invariance for simplicity, time-reversal symmetry is microscopically implemented via detailed balance: if at time $t$ the microstate of the system is $\boldsymbol{\rho}^\prime$, and at time $t=0$ the microstate is $\boldsymbol{\rho}_0$, then 
\begin{eqnarray}\label{detbal}
        \mathrm{P}(\boldsymbol{\rho}^\prime, t | \boldsymbol{\rho}_0, 0) P_{\mathrm{eq}}(\boldsymbol{\rho}_0) =  \mathrm{P}(\boldsymbol{\rho}_0, t | \boldsymbol{\rho}^\prime, 0) P_{\mathrm{eq}}(\boldsymbol{\rho}^\prime)
\end{eqnarray}
Here $\mathrm{P}(\cdots)$ denotes the transition probability, which can be calculated via path integral: \cite{msr} 
\begin{eqnarray}
        \mathrm{P}(\boldsymbol{\rho}^\prime, t | \boldsymbol{\rho}_0, 0) = \int\limits_{\boldsymbol{\rho}(0) = \boldsymbol{\rho_0},\boldsymbol{\rho}(t) = \boldsymbol{\rho}^\prime}  \mathrm{D} \boldsymbol{\rho} \mathrm{D} \boldsymbol{\pi}\ \mathrm{e}^{\mathrm{i} \int \mathrm{d} t \mathcal{L}}.
\end{eqnarray}
Observe that the transformation (\ref{KMS}) is accompanied with $t\rightarrow -t$, which flips the two boundary conditions in the path integral.  Combining (\ref{Sshift}) with (\ref{ss}) we obtain (\ref{detbal}).  Alternatively, demanding (\ref{detbal}) and the invariance of $H$ under detailed balance, we are led to demand (\ref{KMS}).  We deduce that (\ref{KMS}) is true, \emph{independent of whether or not $\epsilon$ is small.} 

So far, our discussion has focused on theories with Gaussian noise, which are described by a quadratic Hamiltonian $H(\pi,\rho)$.   However, it is straightforward to consider higher order Hamiltonians from the EFT perspective.  What is highly nontrivial is to convert the action $S[\pi,\rho]$ back to the Fokker-Planck equation, once we consider nonlinearities in $\pi$.  Remarkably, the KMS-like symmetry implies that (\ref{Hamiltonian}) holds exactly, to all orders in $\epsilon$, since $H(\boldsymbol{\pi}=\mathbf{0}) = 0$.
In Appendix \ref{app:stat}, we give the generalization of (\ref{Hamiltonian}) to non-perturbatively large noise without time-reversal. 

Eq. (\ref{KMS}) extends to situations where detailed balance is broken. Let us decompose $F_x=F_x^{(\text e)}+F_x^{(\text o)}$, where $F_x^{(\text o)}$ satisfies $\int_x F_x^{(\text o)}\mu_x=0$. It is easy to verify that (\ref{KMS}) still holds if, instead of (\ref{Sshift}), we have
\begin{eqnarray}\label{Sshiftb} S\to S^*+\text i\Delta\Phi/\epsilon,\end{eqnarray}
where $S^*$ is the original action with $F_x^{(\text o)}\to -F_x^{(\text o)}$, and where $F_x^{(\text e)}$ obeys (\ref{Hamiltonian}).  $F_x^{(\text o)}$ is unrelated to the noise $Q_{xx'}$, and correspond to time-reversal breaking terms (hence the sign flip in (\ref{Sshiftb})) that are not dissipative.  In hydrodynamics such terms can arise from quantum anomalies \cite{Vilenkin:1980fu,Son:2009tf}, Hall transport \cite{Avron:1995fg}, and more general situations when boost invariance is broken \cite{deBoer:2017abi}. We conclude that any hydrodynamic theory for $\rho$, with a stationary homogeneous distribution, satisfies symmetry (\ref{Sshiftb}) at leading orders in the derivative expansion (see Appendix \ref{app:kms}).

\section{Fracton fluids}
We now begin to classify the new universality classes of fracton hydrodynamics with or without P or T symmetry. Here we will systematically discuss systems with only three kinds of multipole charge conservation: monopole, dipole and quadrupole conservation, but our framework can be easily generalized to other systems.  At least for multipole conserving theories, it appears that all of the peculiar possible phenomena can be found already within one of these three theories.

We start by writing down all possible leading-order terms in Hamiltonian; namely, we will consider at most quadratic terms in $\pi$, and keep as few derivatives and nonlinearities in $\rho$ or $\mu$ as possible.

\textbf{Charge conserving:} the action is invariant under the transformation
    \begin{eqnarray}
		\pi \rightarrow \pi + c(t),
	\end{eqnarray}
so the Hamiltonian should be function of ${\partial_{x} \pi}$ or higher order derivative terms: 
\begin{align} \label{Hc}
    H = A(\rho)\partial_x \pi - \sigma(\rho)\partial_x \mu \partial_x \pi - \mathrm{i}\epsilon Q(\rho)(\partial_x\pi)^2 + \cdots ,
\end{align}

 \textbf{Dipole conserving:} $H$ should consist only of $\partial_x^2 \pi$ or higher order terms: \begin{align} \label{Hd}
        H =  A(\rho)\partial_x^2 \pi + \partial_x B(\rho) \partial_x^2 \pi - \sigma(\rho)\partial_x^2 \mu \partial_x^2 \pi \notag \\
        -  \mathrm{i}\epsilon Q(\rho)\left(\partial_x^2\pi\right)^2 + \cdots ,
\end{align}

  \textbf{Quadrupole conserving:} $H$ should consist only of $\partial_x^3 \pi$ or higher order terms: 
\begin{align} \label{Hq}
        H =  A(\rho)\partial_x^3 \pi + \partial_x B(\rho) \partial_x^3 \pi+ \partial_x^2 C(\rho) \partial_x^3 \pi \notag \\
        - \sigma(\rho)\partial_x^3 \mu\partial_x^3 \pi - \mathrm{i}\epsilon Q(\rho)\left(\partial_x^3\pi\right)^2 + \cdots ,
\end{align}

In the above equations, ${A(\rho)}$, ${B(\rho)}$, ${C(\rho)}$, ${\sigma(\rho)}$, and $Q(\rho)$ are (as of yet) undetermined functions of ${\rho}$, which don't include any derivatives. Combining all other the constraints we imposed to the system, (\ref{Hamiltonian}), (\ref{cond1}) and (\ref{KMS}), we list all possible forms of the undetermined functions in Table \ref{thetable}. From the table, we see that with or without P or T, the leading order dissipative terms ${\sigma(\rho)}$ are always the same and are fixed by the conditions (\ref{Hamiltonian}) and (\ref{cond1}). This is the fluctuation-dissipation theorem \cite{kwon}. Second, when the systems have PT symmetry or neither, there always exists a nonzero leading order term, which is dissipationless, and can lead to instabilities.  In the charge-conserving case, the endpoint of this instability is the KPZ fixed point \cite{kpz,Spohn_2014,PhysRevE.90.012124,gloriosoprl}; in higher dimensions, we have found a new generalization of KPZ.

\begin{table}[t]

\begin{tabular}{ |c|c|cccc| } \hline
 conservation & symmetry & $A(\rho)$ & $B(\rho)$ & $C(\rho)$ & $\sigma(\rho)$ \\ 
 \hline
   & T or P & 0 &   &   & $Q$ \\ 
 monopole & PT & $f(\mu)$ & / & / & $Q$ \\ 
   & None & $f(\mu)$ &   &   & $Q$ \\ 
 \hline
   & T or P & 0 & 0 &   & $Q$ \\ 
 dipole & PT & 0 & $\mu$ & / & $Q$ \\ 
   & None & 0 & $\mu$ &   & $Q$ \\ 
 \hline
   & T or P & 0 & 0 & 0 & $Q$ \\ 
 quadrupole & PT & $\mu$ & 0 & $\mu$ & $Q$ \\ 
   & None & $\mu$ & 0 & $\mu$ & $Q$ \\ 
 \hline
\end{tabular}
\caption{Leading order terms in $H$ for a fracton fluid. ${f(\mu)}$ represents an arbitrary function of ${\mu}$.}
\label{thetable}
\end{table}

To estimate the critical dimensions for these fixed points, we assume that the charge susceptibility $\int \mathrm{d}^d x \langle\rho^2(x)\rangle$ is finite, which implies the scaling $\rho\sim L^{-\frac d2}$, where $L$ is the system size. In the charge conserving case, the leading nonlinearity in the current is $J_x=A(\rho)\sim \rho^2\sim L^{-d}$, while the leading dissipative term is Fick's law $J_x=-\sigma \p_x\mu\sim \p_x\rho\sim L^{-1-\frac d2}$. We see that, as $L\to 0$, the nonlinearity dominates over the dissipative term below $d=2$. Taking $B(\rho)$ to be the leading nonlinearity in the dipole conserving case (see Table \ref{thetable}), a similar reasoning gives $d=2$ as critical dimension, while in the quadrupole conserving case, with the leading nonlinearity begin $B(\rho)$, $d=6$.  For $n$-pole conserving systems in general, we find upper critical dimension $d=2(1+n)$ if $n$ is even, and $d=2n$ if $n$ is odd.  Hence for sufficiently large $n$, the upper critical dimension for hydrodynamics can be arbitrarily large.

We can also answer the question we posed at the beginning of the letter, under (\ref{jxx}). We cannot write ${J_{xx} = -D' \rho + \cdots}$, because the dissipationless part of the dispersion relation can change if we break T or P symmetry, but the leading order dissipative terms in the systems (within linear response) do not change.  This follows from the requirement of stationarity,  (\ref{Hamiltonian}).

	\section{Numerical simulations}
We now present large-scale simulations of classical Markov chains in one-dimensional lattice models with quadrupole conservation, and with or without time-reversal symmetry.  The time-reversal symmetric chain is constructed generalizing \cite{knap2020,iaconis2021}: we allow charges of value $q_x = 0,\pm 1,\ldots, \pm 4$ to exist on each of $L$ sites of a 1d lattice, with periodic boundary conditions;  at each time step, we act with ``gates" on each $q$-tuple of adjacent sites, and replace the configuration of charges present with another one with identical charge, dipole and quadrupole moment.  We have taken $q=6$ in our simulations to ensure the dynamics does not get frozen \cite{shattering,2020PollmannFragmentation} and that the late-time physics is captured by hydrodynamics.  

\begin{figure}[t]
\centering
\includegraphics[scale=0.8]{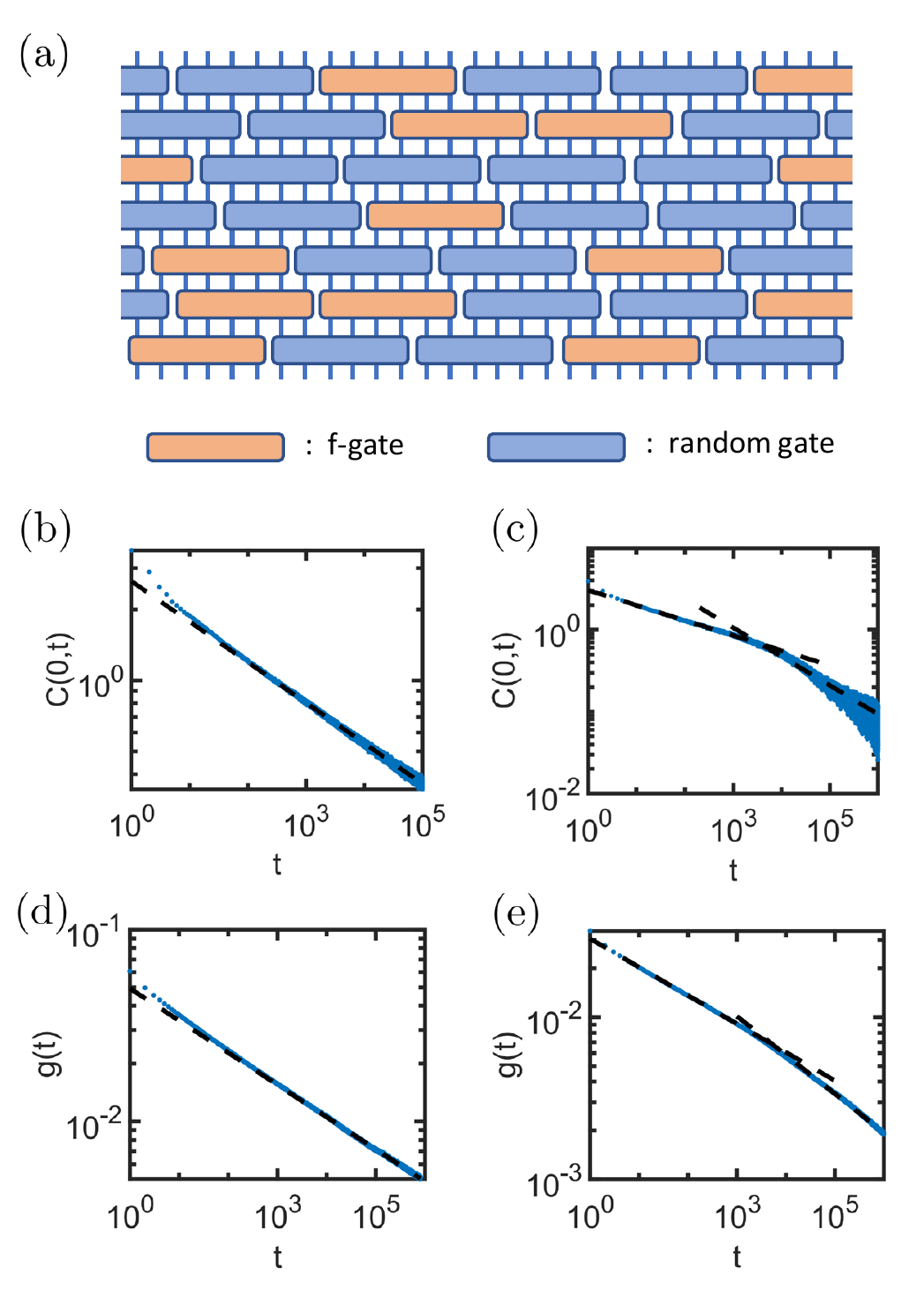}
\caption{(a) Sketch of the Markov chains we simulate. At each time step, we act with f-gates (that break time-reversal symmetry) with probability $p=0.1$, and random gates with $1-p=0.9$, on blocks of size 6. (b) $C(0,t)$ with time-reversal symmetry. The dashed line ${\sim t^{-1/6}}$. (c) $C(0,t)$ without time-reversal symmetry. The dashed line at early-time ${\sim t^{-1/6}}$. The dashed line at late-time ${\sim t^{-1/3}}$. (d) $g(t)$ with time-reversal symmetry. The dashed line ${\sim t^{-1/6}}$. (e) $g(t)$ without time-reversal symmetry. The dashed line ${\sim t^{-1/4}}$. All plots use $s=5\times 10^4$}
\label{longfigure}
\end{figure}

We analyze the correlator \begin{eqnarray}
        C(x,t) = \langle q_{x+y}(t+s)q_y(s)\rangle_{y},
\end{eqnarray}
with the average taken over position $y$, and random realizations of the gates and initial conditions; the correlator is insensitive to the value of $s \lesssim 10^5$.  With time-reversal symmetry, by dimensional analysis we know that $C(0,t) \sim t^{-1/z}$ with $z=6$; as in \cite{knap2020,iaconis2021} we can confirm this scaling readily in numerics: see Figure \ref{longfigure}.

Now let us sketch how we break time-reversal symmetry: details are found in Appendix \ref{app:numerics}.  If we only had charge conservation, then we could break time-reversal symmetry by simply hopping a unit of charge to the right neighbor with some finite probability at the end of each round of random gates.  Importantly, this rule does not modify the fact that the uniform distribution (taken over all many-body configurations in each fixed charge sector) is the stationary distribution of the classical Markov chain: thus, we can readily numerically evaluate $C(x,t)$ by sampling uniformly random initial conditions.  To generalize this idea to $n$-pole conserving models, first observe that the charge conserving chain can be understood as operating by always trying to increase the local dipole moment.  We modify this picture by finding gates which try to increase the $(n+1)$-pole moment in each charge sector, yet do so disturbing the uniform distribution as little as possible.  While we could not find a Markov chain which provably has a uniform many-body stationary distribution once $n>0$, the chains which we did find exhibit behaviors which are consistent with our qualitative expectations:  namely, breaking P and T leads to a dissipationless ``drift" term.  When $n>0$, the drift term will lead to a power-law decay: \begin{eqnarray}
        C(0,t) \sim \left\lbrace \begin{array}{ll} t^{-1/(n+1)} &\ n \text{ even} \\ t^{-1/(n+2)} &\ n \text{ odd} \end{array}\right..
\end{eqnarray}  
To estimate the dynamical critical exponent, we must discard this added drift, so we calculate \begin{eqnarray}
        g(t) \equiv \int \mathrm{d}x \; C(x,t)^2 \sim t^{-1/z}.
\end{eqnarray}
as our estimate for the dynamical exponent $z$.  Intuitively this correlator will capture the ``width" of an initial charge distribution at time $t$.  Figure \ref{longfigure} shows that after an initial transient period of $z=6$ scaling, at sufficiently late times the chain exhibits anomalous scaling with $z\approx 4$.  This is consistent with the existence of a new dynamical universality class, whose upper critical dimension will be $d=6$.


		\section{Outlook}
		In this letter we have described the systematic construction of nonlinear fluctuating hydrodynamics without time-reversal symmetry.  Our construction is valid with or without a well-defined temperature, generalizing recent field theories of hydrodynamics \cite{Crossley:2015evo,Haehl:2015foa,Jensen:2017kzi} to a broad range of theories which cannot be coupled to a spacetime metric.  A non-trivial example of this is to multipole-conserving theories, where we have shown that hydrodynamics can break down in PT-symmetric models; the late time physics is described by exotic dynamical universality classes.  We hope to report on additional applications of our formalism in the near future.

\section*{Acknowledgements}
This work was supported in part by the Alfred P. Sloan Foundation through Grant FG-2020-13795 (AL) and FG-2020-13615 (PG), the National Science Foundation under CAREER Grant DMR-2145544 (JG, AL), the Department of Energy through Award DE-SC0019380 (PG), the Simons Foundation through Award No. 620869 (PG), and through the Gordon and Betty Moore Foundation's EPiQS Initiative via Grant GBMF10279 (JG, AL).

\begin{appendix}
\section{Stationarity condition for non-perturbative noise}\label{app:stat}
Here we derive the requirement on $H$ that must be satisfied if the stationary distribution of the nonlinear fluctuating hydrodynamics is $\mathrm{e}^{-\Phi}$.   Observe that the path integral must obey \begin{align}
    \mathrm{e}^{-\Phi(\rho^\prime)} = \int \mathrm{D}\rho_0 \mathrm{e}^{-\Phi(\rho_0)} \int\limits_{\rho(\mathrm{d}t) = \rho^\prime, \rho(0) = \rho_0} \mathrm{D}\pi \mathrm{D}\rho \mathrm{e}^{\mathrm{i}S};
\end{align}
namely, if we evolve the stationary distribution for time $\mathrm{d}t$ it does not change.  Now, letting $\pi_0 = \pi(t=0)$ and evaluating the path integral using infinitesimal time steps,
\begin{equation}
    S \approx \int \mathrm{d}^dx \; \left[\pi_0 (\rho^\prime - \rho_0) - \mathrm{d}t \times H(\pi_0, \rho_0) \right],
\end{equation}
and therefore 
\begin{align}
        \mathrm{e}^{-\Phi(\rho^\prime)} &\approx \int \mathrm{D}\rho_0 \mathrm{e}^{-\Phi(\rho_0)} \int \mathrm{D}\pi_0 \mathrm{e}^{\mathrm{i}\int \mathrm{d}^dx \pi_0(\rho^\prime - \rho_0)} \notag \\
        &\left(1 - \mathrm{i}\times \mathrm{d}t \times\int \text{d}^d x\, H(\pi_0, \rho_0) + \mathrm{O}(\mathrm{d}t^2)\right) \notag \\
        &\approx \mathrm{e}^{-\Phi(\rho^\prime)} - \mathrm{i}\times \mathrm{d}t \times \int \mathrm{D}\rho_0 \mathrm{e}^{-\Phi(\rho_0)} \notag \\
        &\int \mathrm{D}\pi_0 \mathrm{e}^{\mathrm{i}\int \mathrm{d}^dx \pi_0(\rho^\prime - \rho_0)}\int \text d ^d x\, H(\pi_0, \rho_0). 
\end{align}
We deduce that the second term above must equal 0.  Now, observe that the last term in the last equation above is effectively a Fourier transform from $\pi_0$ to $\rho^\prime$, which we can undo: \begin{align}
    0  = \int \mathrm{D}\rho_0 \; \mathrm{e}^{-\Phi(\rho_0) - \mathrm{i}\int \mathrm{d}^dx \pi_0 \rho_0} \int \text d ^d x\, H(\pi_0, \rho_0). 
\end{align}
This is the generic requirement on $H$ in order to have stationarity. It is transparent to implement in the limit where $\Phi$ is sharply peaked, as the $\rho_0$ integral may be done via saddle point.  The saddlepoint equation gives \begin{align}
    \mu(\rho_0) = \mathrm{i}\pi_0,
\end{align}
and therefore the criterion that when noise is weak, (\ref{Hamiltonian}) must hold.

\section{KMS invariance of hydrodynamics for general homogeneous stationary states}\label{app:kms}
In this appendix, we show that hydrodynamic fluctuations around a locally homogeneous stationary state always satisfies KMS invariance (\ref{Sshiftb}) at leading order, irrespective of whether such stationary state is thermal or not, and independently of the existence of microscopic time reversal.

First, we observe that the linearized dynamics around a stationary distribution always satisfies KMS invariance. Indeed, let us assume that (\ref{langevin})-(\ref{noise1}) describe a linear stochastic process, i.e. $F_x(\boldsymbol\rho)=(F^{(\text e)}_{xx'}+F^{(\text o)}_{xx'})\rho_{x'}$, where $F^{(\text e)}_{xx'}$, $F^{(\text o)}_{xx'}$ and $Q_{xx'}$ are constant, and $F^{(\text e)}_{xx'}$, $F^{(\text o)}_{xx'}$ are defined below (\ref{Sshiftb}). Plugging these together with $\Phi(\boldsymbol\rho)=\frac 12 U_{xx'}\rho_x\rho_{x'}$  into eq. (\ref{fokkerplanck}) we find
\begin{eqnarray}
F^{(\text e)}_{xy}U_{xz}+F^{(\text e)}_{xz}U_{xy}=Q_{xw}U_{xy}U_{wz},
\end{eqnarray}
which is precisely the relation one finds by imposing KMS symmetry (\ref{Sshiftb}). Here, however, this relation is simply a consequence of stationarity, without imposing any stronger condition. In general, this statement holds only for linear perturbations around a stationary state \cite{kwon,pingao}. In particular, this means that the Hamiltonian (\ref{lagrangiandensity}) will \emph{always} satisfy KMS invariance at quadratic order in amplitude expansion so long as it describes the dynamics around a stationary state.

Let us now consider the case in which $\boldsymbol\rho$ is a conserved quantity. The Hamiltonian in this case has the form (\ref{Hc}), which we expand in amplitude $\rho_x=\bar\rho+\delta\rho_x$ around the background value $\bar\rho$ up to quadratic order in amplitude perturbation.
Since $\boldsymbol\rho$ is a conserved quantity, this Hamiltonian describes the linearized dynamics around a homogeneous stationary state with background density $\bar\rho$, and must therefore satisfy KMS symmetry. Varying over the values of $\bar \rho$, we then see that this will constrain nonlinear terms as well, as far as they contribute to the quadratic Hamiltonian. The term with lowest number of derivatives that is not constrained by this procedure (and which is thus not, \emph{a priori}, KMS invariant) is $H\sim C_{ijk}\p_i\mu\p_j\mu\p_k\pi$,
which is highly suppressed as we are interested in the long-wavelength dynamics, and can generally be neglected. A similar discussion can be done in higher dimensions and with conserved higher multipoles. In particular, all allowed terms listed in Table \ref{thetable} satisfy the KMS symmetry (\ref{Sshiftb}).

\section{Details on the classical Markov chains}\label{app:numerics}

Here we show the details of how we break time reversal symmetry in our Markov chain simulations, focusing on quadrupole-conserving systems as an example.   Our goal is to break time reversal symmetry while keeping the many-body stationary distribution of the Markov chain uniform. 

Before discussing microscopic update rules, it is helpful to consider what we hope to find.  We expect that the ${A(\rho)\partial_x^3 \pi}$ term in (\ref{Hq}) -- namely, the $A(\rho)$ term in the quadrupole current $J_{xxx}$ -- is not negligible. The consequence of a non-vanishing ${A(\rho)\partial_x^3 \pi}$ can be seen from the time derivative of the octopole moment:
\begin{align}
    \partial_{t} \int \mathrm{d}x\ x^{3} \rho = -6\int \mathrm{d}x\ A(\rho),
\end{align}
which means when we act the gates and replace the blocks of charges, we want the octopole moment of the block to increase more if the block has positive net charge than when the block has negative net charge.  (If the octopole moment of the block tends to increase regardless of $\rho$, that will contribute a constant term in $A$ which drops out of equations of motion!)

With time-reversal symmetry, we create a dictionary of all possible 6-site configurations with fixed charge, dipole, and quadrupole charges \cite{knap2020,iaconis2021}.   Our time-reversal and parity-preserving chain consists of choosing groupings of 6 adjacent sites at random (we do so in parallel across the entire chain, so each site gets updated once per time step), and replacing each configuration of charges within a grouping with another one, drawn from the dictionary, of the same charges, uniformly at random. We call this applying a ``gate", as in the literature on random quantum circuits.   It is believed that the unique stationary distribution of this Markov chain, at fixed charge, dipole, and quadrupole charge, is uniform -- at least close to zero charge density.

We time-reversal breaking by, with some probability $p$, replacing the gates above with a T-breaking gate, which we define as follows.   Consider some function $f_x$ obeying 
\begin{equation}
    \sum_{x} f_{x} = 0
\end{equation}
for any state.  A typical example of this is to set \begin{eqnarray}
        f_x = \sum_{j=0}^5a_j q_{j+x}
\end{eqnarray}
where $q_x = 0, \pm 1,\ldots,\pm 4$ denotes the charge on site $x$, and to demand 
\begin{eqnarray} 
        \sum_{j=0}^5 a_j = 0.
\end{eqnarray}
To implement a T-breaking gate, we first calculate ${f_x}$ for each block. If the net charge of the block is negative and ${f_x > 0}$, or if the net charge of the block is positive and ${f_x < 0}$, with a probability ${ \propto \abs{f_x}}$, we replace the block with a block that has ${-f_x}$ and identical multipole moments (charge, dipole and quadrupole moments). Otherwise, the block stays the same. Under this rule, for each replacement, the probability of a certain chain being changed to another chain will be
\begin{equation}
    P_{\rm out} \propto \sum f_{x}^{-}\ \mathrm{\Theta}(f_{x}^{-}) - f_{x}^{+}\ \mathrm{\Theta}(-f_{x}^{+}),
\end{equation}
where ${f_{x}^{-}}$ denotes ${f_x} \mathrm{\Theta}(-q_x - \cdots - q_{x+5})$ -- i.e. $f_x$ restricting to blocks with negative net charges -- while ${f_{x}^{+}}$ corresponds to blocks with positive net charge.  Here ${\mathrm{\Theta}(x)}$ is the unit step function. Observe that if the stationary distribution of the Markov chain was uniform (i.e. up to global conservation laws $P_{\mathrm{eq}}(q_1,\ldots, q_L) = c$ for some constant $c$) the probability of a certain microstate $\boldsymbol{q}$ being allowed to transition into another state is
\begin{equation}
    P_{\rm in}(\boldsymbol{q}) \propto \sum f_{x}^{+}\ \mathrm{\Theta}(f_{x}^{+}) - f_{x}^{-}\ \mathrm{\Theta}(-f_{x}^{-}).
\end{equation}
Now we have
\begin{equation}
    P_{\rm out} - P_{\rm in} \propto \sum f_{x}^{-} - f_{x}^{+}. \label{eq:Pout}
\end{equation}
If for large systems,
\begin{equation}
    \sum f_{x}^{-} \approx \sum f_{x}^{+} \approx 0,
\end{equation}
we expect that ${P_{\rm out} \approx P_{\rm in}}$, so that the uniform distribution will be approximately stationary (and thus it is easy to sample by preparing the chain in a microstate chosen uniformly at random).

To make the effect of the ${A(\rho)\partial_x^3 \pi}$ term manifest, we desire that blocks with higher ${f_x}$ should have higher (or lower) octopole moments.  We do so by choosing ${a = [-1,\  1,\  0,\  0,\  -1,\  1]}$.  Unfortunately, it turns out that only 80\% of configurations $(q_x,\ldots, q_{x+5})$, with a given $f_x$, have $\ge 1$ ``partner" with the same charge, dipole and quadrupole moments but $-f_x$.  For this reason, (\ref{eq:Pout}) does not exactly hold in our chain.



\begin{figure}[t]
\centering
\includegraphics[scale=0.28]{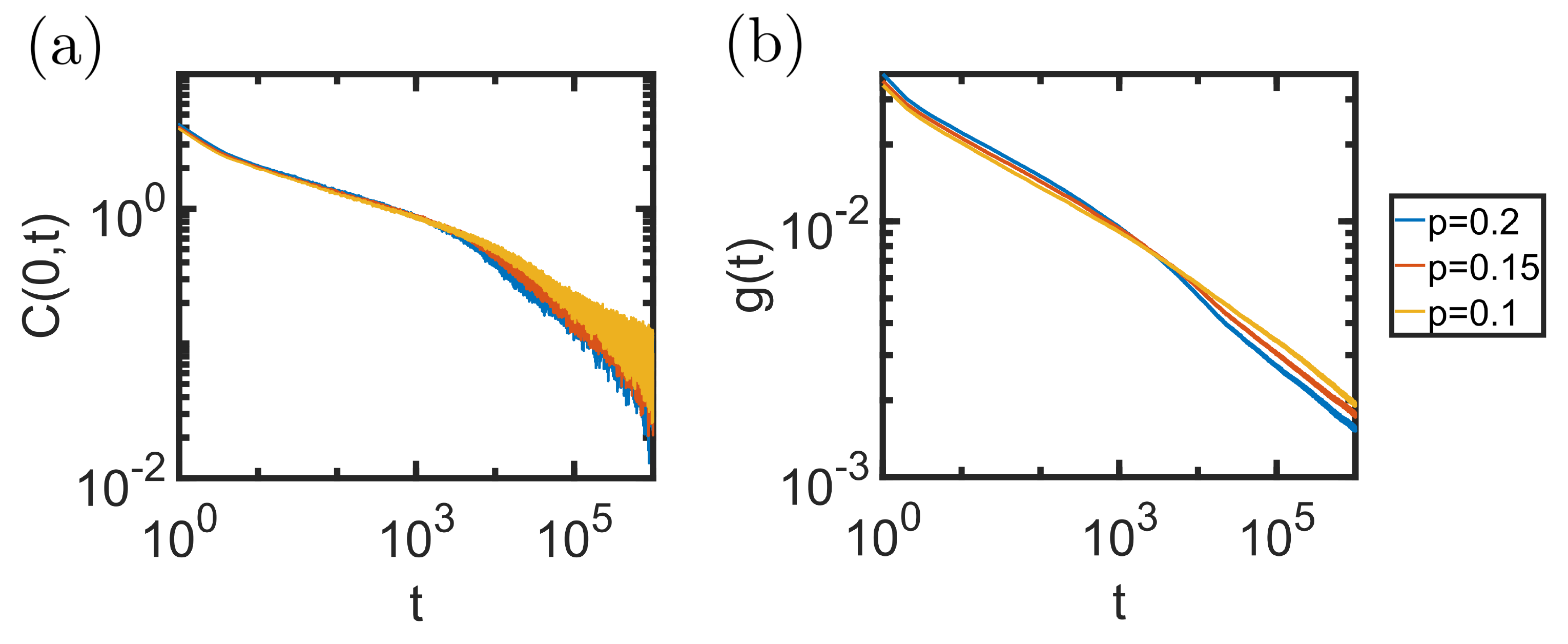}
\caption{The data of $C(0,t)$ and $g(t)$ with different probabilities of applying f-gates. The early time behaviors and late time behaviors of the correlation functions with different $p$s are almost the same. The fluctuations in the orange plot in (a) are due to sampling noise.}
\label{longfigure2}
\end{figure}

We name these T-breaking operations ``f-gates''. Note that the system will freeze up if we only apply ``f-gates", so it is important to apply more of the random and thermalizing gates vs. f-gates (i.e. $p\ll 1$).   Nevertheless, we find that for $p\sim 0.1$, the dynamics appears to be thermalizing for $>10^6$ time steps, and that (after allowing the chain to thermalize for a similar number of initial steps), the two-point function $C(x,t)$ exhibits behaviors consistent with our EFT predictions in the main text.  We have further checked that -- although we do not know the exact steady-state distribution of these chains -- equal time correlation functions of the charge density are essentially constant in time (to 1 part in $10^4$),  providing evidence that any drift in the probability distribution of the chain with time is not strong enough to explaain the anomalous scaling we observe; rather it seems more likely to be due to the breakdown of subdiffusive hydrodynamics.

\end{appendix}

\bibliography{thebib}

\end{document}